\documentclass[conference]{IEEEtran}
\usepackage{graphicx}
\ifCLASSINFOpdf
\else
\fi
\begin{document}
	
\IEEEoverridecommandlockouts
\IEEEpubid{\makebox[\columnwidth]{978-1-5090-6505-9/17/\$31.00 \copyright 2017 IEEE \hfill } 
\hspace{\columnsep}\makebox[\columnwidth]{\hfill }}
%
\title{Towards Modern Inclusive Factories:\\A Methodology for the Development \\ of Smart Adaptive Human-Machine Interfaces}

\author{\IEEEauthorblockN{Valeria Villani\IEEEauthorrefmark{1}, Lorenzo Sabattini\IEEEauthorrefmark{1}, Julia N. Czerniak\IEEEauthorrefmark{2},\\ Alexander Mertens\IEEEauthorrefmark{2}, Birgit Vogel-Heuser\IEEEauthorrefmark{3} and Cesare Fantuzzi\IEEEauthorrefmark{1}}
\IEEEauthorblockA{\IEEEauthorrefmark{1}Department of Sciences and Methods for Engineering (DISMI)\\University of Modena and Reggio Emilia, Reggio Emilia, Italy\\
Email: \{valeria.villani, lorenzo.sabattini, cesare.fantuzzi\}@unimore.it}
\IEEEauthorblockA{\IEEEauthorrefmark{2}Institute of Industrial Engineering and Ergonomics\\RWTH Aachen University, Aachen, Germany\\
Email: \{j.czerniak, a.mertens\}@iaw.rwth-aachen.de}
\IEEEauthorblockA{\IEEEauthorrefmark{3}Institute of Automation and Information Systems,\\ Technical University of Munich, Munich, Germany\\
Email: vogel-heuser@ais.mw.tum.de}
}


%


\maketitle

\begin{abstract}
Modern manufacturing systems typically require high degrees of flexibility, in terms of ability to customize the production lines to the constantly changing market requests. For this purpose, manufacturing systems are required to be able to cope with changes in the types of products, and in the size of the production batches. As a consequence, the human-machine interfaces (HMIs) are typically very complex, and include a wide range of possible operational modes and commands. This generally implies an unsustainable cognitive workload for the human operators, in addition to a non-negligible training effort.
To overcome this issue, in this paper we present a methodology for the design of adaptive human-centred HMIs for industrial machines and robots. The proposed approach relies on three pillars: measurement of user's capabilities, adaptation of the information presented in the HMI, and training of the user. The results expected from the application of the proposed methodology are investigated in terms of increased customization and productivity of manufacturing processes, and wider acceptance of automation technologies. The proposed approach has been devised in the framework of the European project INCLUSIVE.
\end{abstract}


%
\IEEEpeerreviewmaketitle

\section{Introduction}
Modern automatic machines and robotic cells in production plants are becoming more and more complex because of higher demands for fast production rate with high quality. Over these basic functions, today's  factories need to allow for higher levels of product customisation and variable requirements. To this end, advanced functions are implemented, such as fault diagnosis and fast recovery, fine-tuning of process parameters to optimize environmental resources, fast reconfiguration of the machine and robot parameters to adapt to production change. 

Despite high levels of automation of machines and robots, humans remain central to manufacturing operations since they take charge of control and supervision of manufacturing activities. Human operators interact with machines and robots by means of user interfaces that are the modern cockpit of any production plant. For example, they set up machine production parameters, identify and solve faults, coordinate machine and robot re-configuration to enable adaptation to product changes. These activities are all performed by means of computerized human-machine interfaces (HMIs) that are inevitably becoming more and more complex, as new functions are implemented by the production system \cite{Sheridan_2002, Nachreiner_2006}.

In this new scenario, human operators experience many difficulties to interact efficiently with the machine; this is particularly true for middle age workers who feel uncomfortable in the interaction with a complex computerized system, even if they have a great experience with the traditional manufacturing processes. On the other hand, complex HMIs linked to complex machine and robot functions create a barrier to young inexperienced or disabled people, who are then unable to effectively manage the production lines.

Such an increasing gap between machine complexity and user capabilities calls for smart and innovative human-centred automation approaches that lead to the determination of adequate levels of automation for optimal flexibility, agility and competitiveness of highly customised production on the one side, and, on the other side, a sustainable effort for \emph{all} workers. Accordingly, novel automation systems should embed HMIs that accommodate to the workers' skills and flexibility needs, by compensating their limitations (e.g. due to age or inexperience) and by taking full advantage of their experience.

Moving along these lines, in this paper we present a methodology for the design of adaptive human-centred HMIs for industrial machines and robots. It consists in enabling user interfaces to measure the user capabilities, experience and cognitive burden and adapt the complexity and information load accordingly.
In particular, according to the proposed methodology, an adaptive user interface for industrial machines can be developed that fully adapts to user's (1) physical status and impairments, (2) cognitive status and mental workload, and (3) experience in the working scenario and in the use of computers. Adaptation concerns visual presentation of information, selection of displayed content, selection of machine functionalities enabled to the user and guidance in the interaction with the process through default recipes and working strategies. Additionally, the interface provides off-line and, more importantly, on-line training to the user in order to increase her/his performance and prevent errors. These solutions aim at improving worker's situation awareness for a more effective, reliable and prompt interaction with the system, thus allowing workers to have a full comprehension of the system behavior and facilitate intervention in dynamic and unforeseen situations.

The final goal is to create an \emph{inclusive} \cite{Abascal_2005, Stephanidis_2001} and flexible working environment for any kind of operator, taking into account multiple cultural background, skills, age and different abilities.
To achieve this, it is needed to reverse the paradigm from the current belief that "the human learns how the machine works" to the future scenario in which "the machine adapts to the human capability" accommodating to her/his own time and features. This is realized by adaptively simplifying the HMI based on the user's features and complementing her/his cognitive capabilities by advanced sensing and the higher precision of machines.
However, this simplification might lead to the increase of the time needed to perform a process function and the reduction of productivity due to limited functionalities enabled to low skilled users. To overcome this issue, a training facility needs to be integrated in the adaptive HMI that embeds a virtual (or augmented) environment to guide and teach the user to evolve her/his capability aiming at a more efficient process, both in terms of time and quality. The approach presented in this paper has been devised within the framework of the European project INCLUSIVE, which seeks to develop smart and adaptive interfaces for inclusive work environment.

The paper is organized as follows. In Section~\ref{sec:state_art} we present a review of the state of the art on adaptive automation systems. In Section~\ref{sec:methodology} the proposed methodology is described, with a special focus on the proposed rules for adaptation in Subsection~\ref{subsec:adaptation_module}. The expected impact of the application of the proposed methodologies is investigated in Section~\ref{sec:expected_results}. Finally, Section~\ref{sec:conclusions} follows with some concluding remarks.


\section{State of the art}\label{sec:state_art}

In human-computer interaction, the interface is what users see and work with to use a device \cite{Sutcliffe_1988}. In industrial scenarios, the HMI takes care of all visualizations and user's interactions with the data coming from technological processes, and thus allows the user to operate the machine, to observe the system status and, if necessary, to intervene in the process. Customarily, HMIs used in industrial process control applications provide no means to control the amount and form of information displayed during operation. While the user is flexible and adaptable, the system is not. Control systems commonly respond in the same way without regard as to whether the flow of information is low or extremely high, or the level of expertise of the user is good or bad \cite{Viano_2000}. As a consequence, the responsibility for the interaction is placed on the user, who has to adapt to processes determined by the technical system. Moreover, the flexibility required to deal with difficult situations must be provided by the operators alone acting under the pressure of unexpected and rapidly changing hazardous situations. This issue is even more severe if we consider that the amount of monitored data that come from modern plant processes keeps increasing and control systems are becoming more and more complex \cite{Flaspoler_2010, Viano_2000, Skripcak_2013}. Therefore, automation results in working methods that demand increase with regard to stamina, time pressure and the pace of work \cite{Flaspoler_2010}. This leads to detrimental effects on workers' health and safety giving rise to occupational diseases, such as stress or musculoskeletal disorders, as well as to occupational accidents \cite{Flaspoler_2010}.

To tackle this issue, context-dependent automation, also known as adaptive automation, has been considered \cite{Parasuraman_2000, Lee_2013}. Context awareness is the ability of programs, applications or computer devices to sense, interpret, respond and act based on the context. Context refers to any information that can be used to characterize the state of an entity, that can be a person, place, or object considered relevant to the interaction between a user and an application, including the user and applications themselves \cite{Dey_2001}. According to this design paradigm, levels of automation need not be fixed at the system design stage, but should be designed to vary depending on situational demands during operational use. In this regard, the distinctive feature of adaptive user interfaces is the possibility to change how the information is presented so that only relevant information is provided to users by including the environment and the user as part of the monitored system through adaptive HMIs.

Adaptive user interfaces have been developed and implemented in different domains, such as automotive \cite{Sharma_2008, Amditis_2006, Garzon_2012}, aeronautics \cite{Inagaki_2000} and smartphone and hand-held devices \cite{Gu_2004}. However, very few partial attempts and preliminary results on the development of adaptive HMIs for complex industrial systems have been reported \cite{Viano_2000, Lee_2013}.  In \cite{Viano_2000} a preliminary conceptual architecture is introduced that allows to defining an HMI that adapts the presentation of information based on the operator responsiveness. In \cite{Lee_2013} different user profiles, such as manager, supervisor and maintenance personnel, are identified and adaptation is limited to present information selectively according to the logged account. Going beyond these preliminary efforts, the methodology we propose in this paper allows for the development of a complete ecosystem of technological innovations that includes the measurement of human capabilities, the adaptation of the user interface and the training of unskilled users.

\section{Proposed methodology}\label{sec:methodology}

\begin{figure}
	\begin{centering}
		\includegraphics[width=\columnwidth]{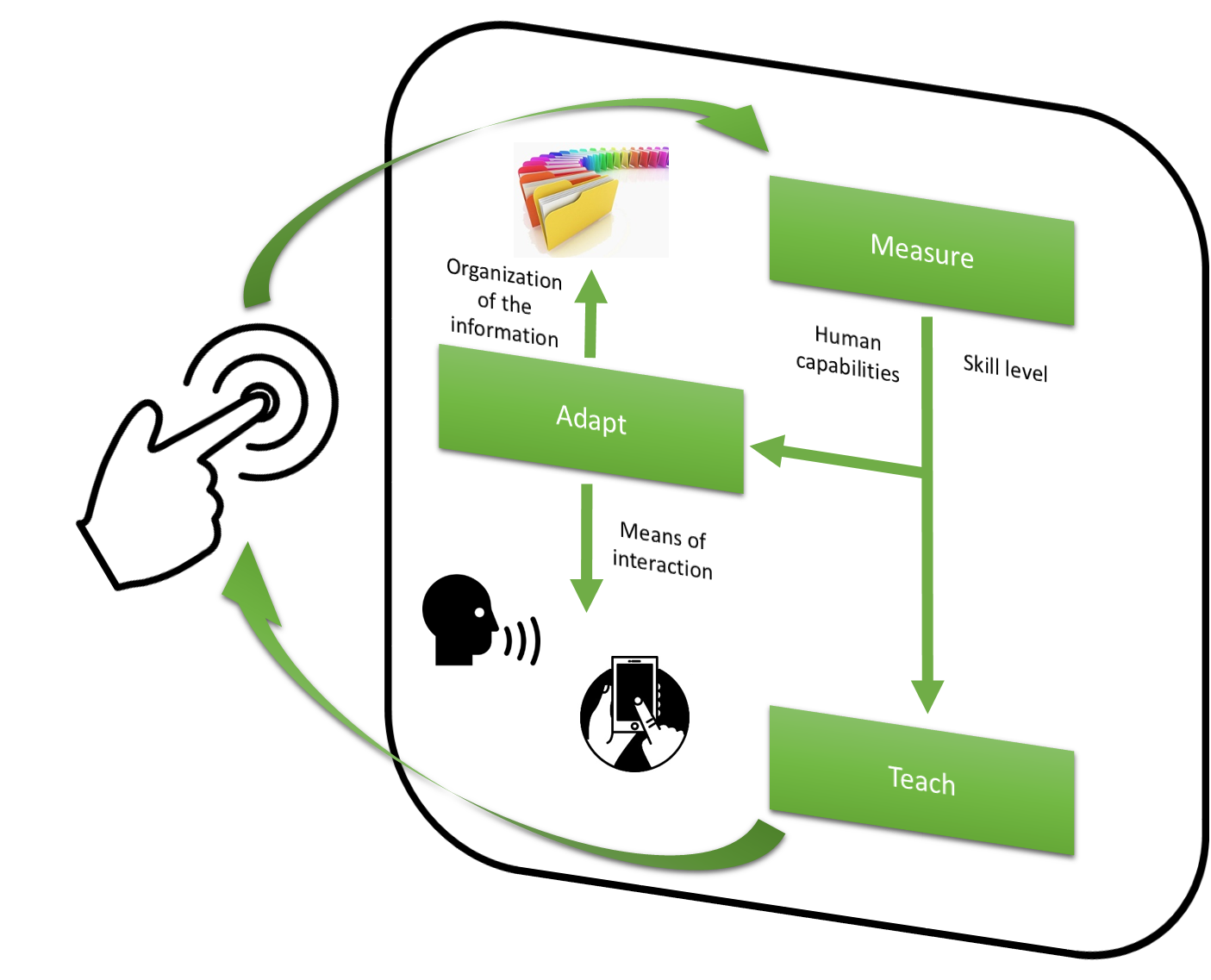}
	\end{centering}
	\caption{\label{fig:scheme}Overview of the proposed approach.}	
\end{figure}

The methodology presented in this paper aims at developing a smart user-machine interface that adapts the information load of the HMI and the automation capability of the machine to the physical, sensorial and cognitive capabilities of workers. The smart interface is based on three main modules, as shown in Figure~\ref{fig:scheme}:
\begin{enumerate}
	\item human capabilities measurement (\emph{Measure}): the smart interface measures the human capability of understanding the logical organization of information and the cognitive burden she/he can sustain (automatic human profiling). The interface identifies also the real skill level of the user analysing how she/he operates in the common working processes (e.g. measuring the time needed to move among different screens of the HMI, measuring the eyes activity in seeking information, etc.);
	\item adapt interfaces to human capabilities (\emph{Adapt}): the smart interface adapts the organization of information (e.g. the complexity of the information presented), the means of interaction (e.g. textual information, only graphics, speech, etc.), and the automation task (normal operation, adaptation to new processes, predictive maintenance, etc.) that are accessible to the user depending on her/his measured capabilities;
	\item teaching and training for unskilled users (\emph{Teach}): the smart interface is used to teach the unskilled users how to interact with the machine. Depending on the skill level of the user and the operation performed by the machine, the interface can train the user by using a step by step procedure, also supported by simulation on a virtual environment. This teaching mode can be on-line or off-line, depending on the level of automation and the criticality of the job operated by the machine or robot. Moreover, in this module, an industrial social network app (Android and iOS) is developed to facilitate the sharing of knowledge among the users about the industrial processes and the machine operational modes.
\end{enumerate}

Since the behaviour of the interface depends on the actual process organization and operational modes, which are specifically related to the particular industrial process under consideration, it is important to establish a general methodological approach that, then, can be specified by building customized HMIs according to applications. Thus, the design of such universal adaptation patterns leads to a core \emph{meta-HMI}, which is general and dialogue-independent from hardware. This \emph{meta-HMI} needs, then, to be customized to the specific application scenarios, functionalities and hardware targets of the use cases.


In the following, the three modules will be presented separately, with a special focus on the adaptation module, which is the core of the system.

\subsection{Measurement of human capabilities}\label{subsec:measurement_module}

The first step towards adaptation is the measurement of the individual capabilities and strain level while fulfilling operative tasks. 
Firstly, the effect of age (changed perception, cognition and motor skills), dyslexia, second-language speaking, disabilities (one handed operation, colour blindness, etc.), missing experience in the context and impaired abilities in acquiring knowledge are measured a priori, at the first involvement of the user with the automatic system. Then, the strain of the operator is continuously assessed in real time. To this end, contactless and body-worn sensors are used to measure several physiological indicators, such as heart rate variability and electrocardiographic activity, galvanic skin response, eye tracking, blink reflex, skin temperature, cerebral electrical activity, and adrenaline/noradrenaline levels \cite{Kulic_2007, Rani_2002}.

\subsection{Adapt interfaces to human capabilities}\label{subsec:adaptation_module}
\begin{figure*}
	\centering
	\includegraphics[width=1.5\columnwidth]{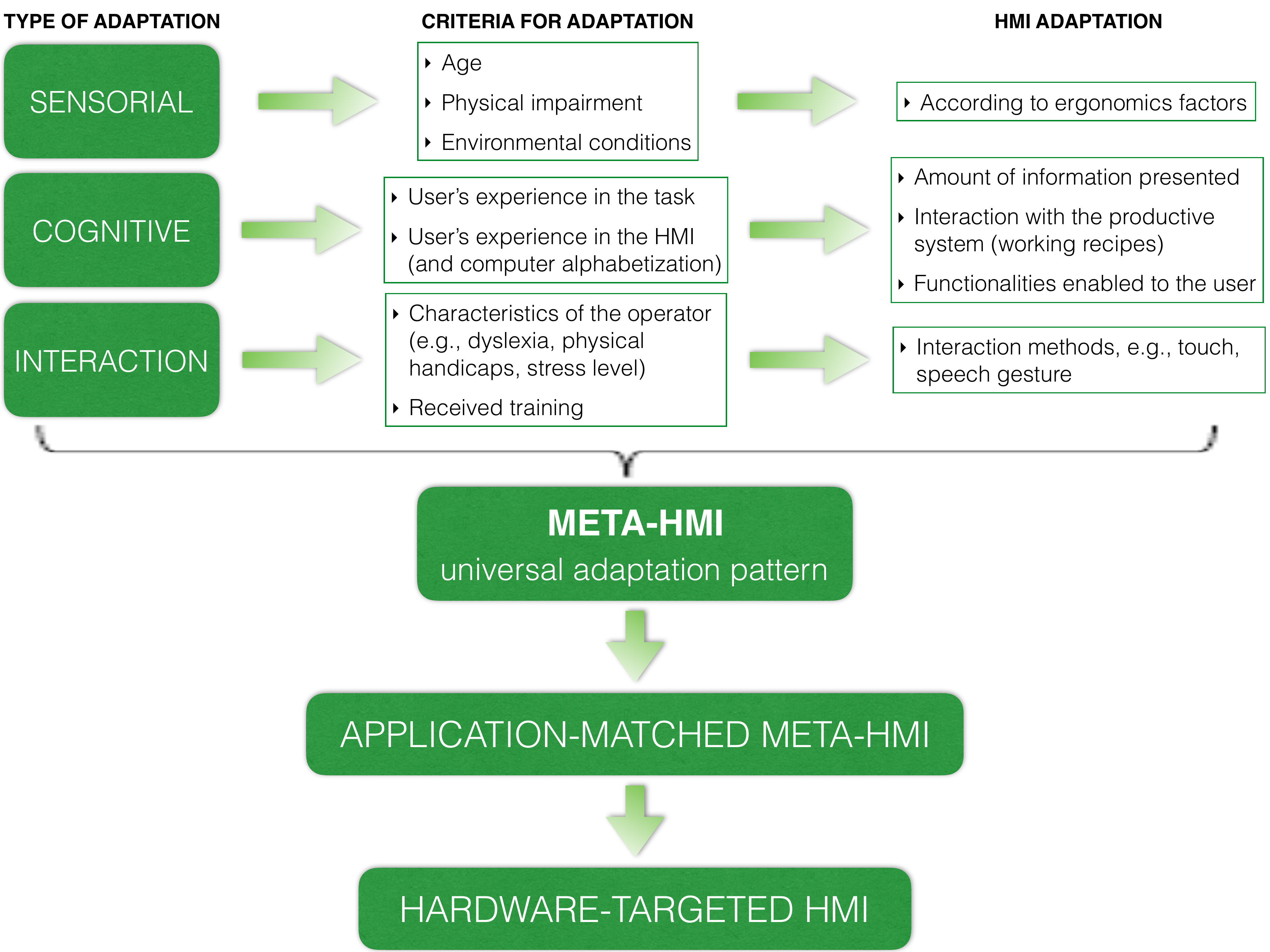}
	\caption{\label{fig:adaptation}Methodological rules for the adaptation module.}
\end{figure*}

Results of the measurement module are directly mapped into a suited degree of adaptation of the interface. Adaptation occurs at  sensorial, cognitive and interaction levels. The proposed approach is summarized in Figure~\ref{fig:adaptation}.

Sensorial adaptation is meant to tackle physical, visual, auditory and dexterity impairments of users. In this regard, the first step towards adaptation consists in meeting user physical impairment mainly by varying the presentation of information, e.g., adapting font size, accompanying icons to short text description, enabling audio input and output. Although such features can be manually enabled/disabled by the user, the optimal configuration is automatically selected by the interface on the basis of user's claims and measurements. Also environmental conditions, such as lighting, noise and use of protective gloves, are considered. A deep analysis based on ergonomics factors \cite{Erg_2006, Czerniak_2017_IFAC} drives the selection of such an optimal configuration.

Cognitive adaptation provides the adequate level of instructions and details in order to not exceed the cognitive capabilities of less experienced workers and increase performance of more experienced workers. It is implemented in terms of: 1) amount of information presented, 2) guided interaction with the productive system, and 3) amount of functionalities enabled to users.
Displayed information is adapted according to two factors: user's experience in the task to accomplish and experience in the use of the HMI (e.g. novel, occasional or habitual user).

In the presence of inexperienced workers, it is considered useful also providing an extended tutorial concerning the description of machine functionalities and/or the use of the HMI (e.g. interaction mode, details on icons, menus, setting of preferences, etc.). This tutorial can be easily accessed while the machine is running and represents a brief and easy to access summary of the teaching module. Novel or occasional workers are provided with a brief description of each button in the interface when the user moves over it with a mouse or finger, depending on the input device. This feature is useful also for elderly users having a long-time experience in the task but limited computer alphabetization, thus being unfamiliar with computer jargon. Conversely, the feature is not enabled in the case of experienced workers, since continuous and ubiquitous explanations are superfluous and would slow down the interaction with the system. Furthermore, depending on worker's familiarity with the interface and the machine, as from findings of the measurement module, information on the machine and the whole process chain are selectively presented to users. Expert workers have access to global view of the process, enriched with information on production rates and trends, levels of input raw material, and due maintenance activities. Additionally, an overview of the plant (or a subpart) is provided by means of an interactive map reporting information about alarms, failures, ongoing tasks and production rates. As regards alarms, experienced users are provided with the detailed list of all active alarms, including those warnings that do not stop production. Conversely, adaptation for less experienced users prescribes a more restricted view of the process, focused on the activity the user is currently performing. Alarms are filtered so that only the most severe ones are shown to the worker, together with a detailed description of causes and actions to take, supported by pictures, videos and technical drawings of the machine. If necessary, secondary alarms are shown at the end of the working day so that the worker can ask for assistance to solve open issues and restore the correct machine status.

Additionally, for users with limited experience in the process or task to be accomplished, the HMI needs to guide the worker in the task. To this end, predefined working recipes (e.g. default values of parameters, working strategies or combination of working parameters) are presented by the interface, covering the range of strategies that can be implemented on the machine. In this way, the cognitive gap between the worker and the system is covered by the interface, and the comprehensibility of the interface is improved.
To support the concept of adaptive automation systems, further adaptation in the HMI is implemented conditioning the accessible functionalities of the machine to user's experience. Indeed, production is adapted to workers' capabilities by disabling most advanced functionalities for workers with limitations due to inexperience or disability, with the goal to decrease mental workload. For example, considering the scenario of an old worker not familiar with the process, exploiting the measurement module, the HMI will automatically disable complex or unusual tasks that require many inputs and are not supported by established recipes.

\subsection{Teaching and training for unskilled users}\label{subsec:training_module}

The last pillar of the proposed approach is an adaptive teaching system that trains the user according to her/his capabilities, identified in the measurement module, and understanding of the working system. Teaching is provided to unskilled users both off-line, before starting a working session, and on-line. The off-line training helps the user to get familiar with the automatic system and learn the task to perform. This is done in a virtual environment replicating the real scenario and working situations. At this stage, the received training is tailored to meet the measured user's capabilities and mental model. Additionally, while the process is ongoing, the user receives additional on-line training that provides guidance in the use of the machine or the robot by means of augmented reality \cite{Kipper_2012}. This module adapts the training level also to the current understanding of the process, assessed, for example, by tacking user's errors and eye.

Additionally, the teaching module hosts an industrial social system, providing a contextual help menu that broadcasts a request for help using a social network media (e.g. app for iOS or Android). In the case of a problem, the operator can contact other qualified experts within the company or, in case that no sufficient help can be provided by them, further experts, e.g. from the plant manufacturer, can be contacted easily. The request is sent to the local workers community or the service level, who has the app installed on her/his mobile device. The HMI augments the message with the detailed context of use of the machine, in order to facilitate the expert to correctly address the problem experienced by the local inexperienced operator. The system routes automatically the request to the users, who qualify at app login as experts in that particular task/machine function. The experts can then respond using a message or voice call to support the unskilled user.

\section{Expected results}\label{sec:expected_results}

The results of the approach proposed in this paper will be measured at the end of the INCLUSIVE project on three different industrial use cases which come from three different market sectors and address different user groups: young, elder artisans, seasonal workmen, people with low level of education, people with certified limited cognitive abilities and physical impairments ranging from mild to severe.
In particular, a use case refers to a large bottling company, with automatic filling and packaging machines, whose employees include people with certified limited cognitive abilities and physical impairments. The second use case is a company producing woodworking machines for artisan shops and small companies. Thus, the final users of the HMI are elder subjects with low education level and computer alphabetization. Finally, the last use case aims at matching a system integrator for robotic applications to a
manufacturing company producing machines for bending metal parts and components, which are currently manually fed mainly because of the variability of the process itself and the lack of skilled personnel, able to manage automatic machines or robots.

\begin{figure*}
	\centering
	\includegraphics[width=1.7\columnwidth]{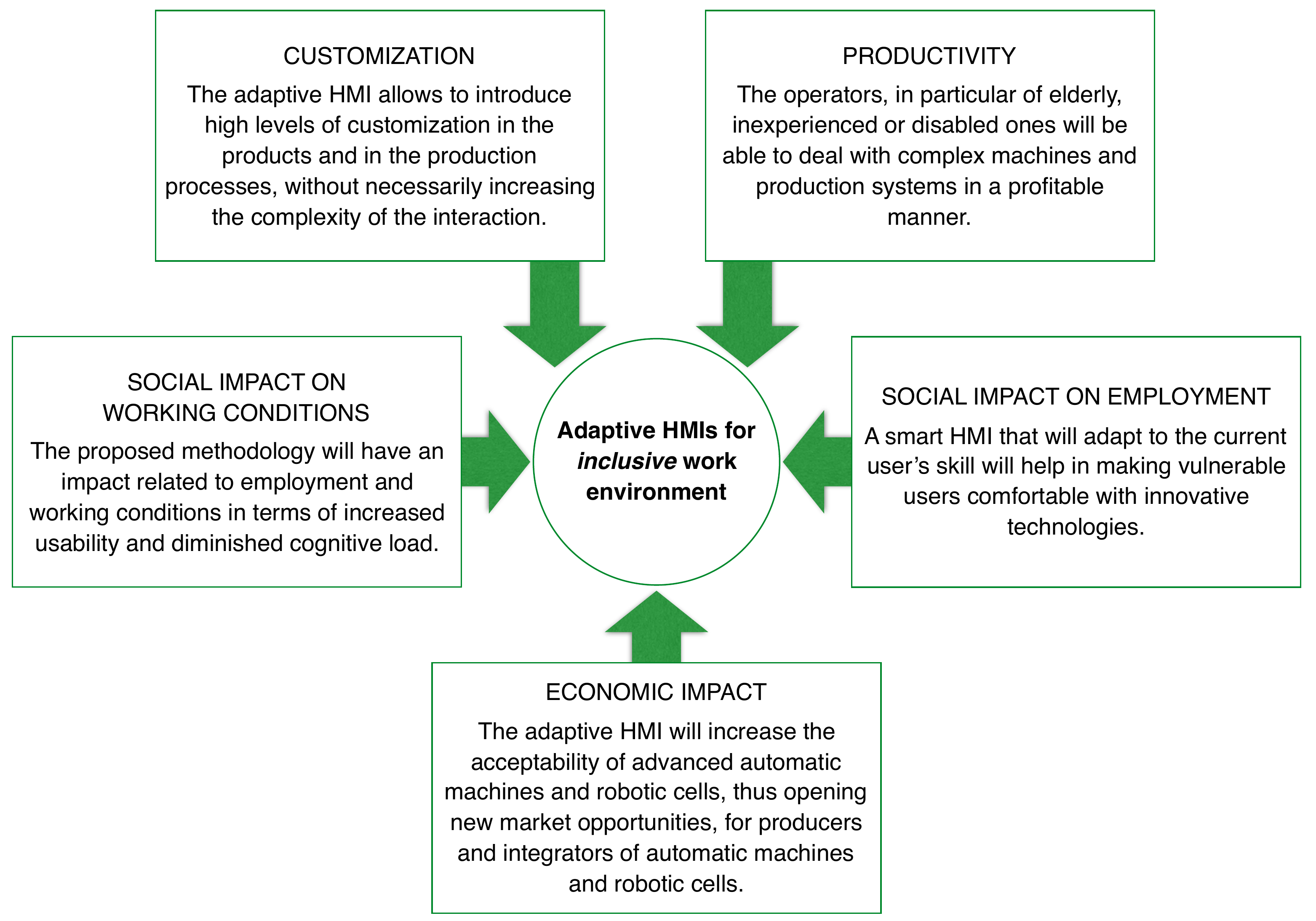}
	\caption{\label{fig:expected_results}Results expected from the proposed methodology.}
\end{figure*}

For the time being, the expected impacts for the increased customization, flexibility optimization of the production and the widest acceptance of automation technologies have been investigated. They are summarized in Fig.~\ref{fig:expected_results}.


\subsection*{Effect on customization of manufacturing processes}

The results of the approach proposed in this paper are expected to have a significant impact on the customization of manufacturing processes, guaranteeing the possibility of introducing significant levels of customization in the products and in the production processes. This will be achieved thanks to the developed smart HMI that will adapt its behaviour at run time, accommodating time-variable needs together with the users' capabilities.

In particular, the HMI developed according to the proposed methodology will make it possible to introduce high levels of customization in manufacturing process machines, while reducing the complexity of the interaction to a sufficiently low level, to enable also non-specialized personnel and operators with disabilities or with low education levels to effectively interact with the machines.

Moreover, as mentioned above for the third use case of the INCLUSIVE project, several manufacturing processes are mainly performed in a manual manner nowadays, due to the high variability of the production batches. Despite the availability of automatic machines able to perform such operations, their potential is often limited by the inability of human operators to interact with such complex systems. In this scenario, the availability of such adaptive user interfaces, which support also on-line and off-line training of operators, allows them to effectively utilize automatic machines.

\subsection*{Effect on productivity of manufacturing processes}
Additionally, we expect that the application of the proposed methodology to the design of adaptive HMIs will have a significant impact on the productivity of the overall manufacturing processes.
Indeed, the performance of the operators, in particular of elderly, inexperienced or disabled ones will be significantly improved since they will be able to deal with complex machines and production systems in a profitable manner.

In particular, the proposed inclusive HMI is expected to allow a significant reduction of the time needed to complete each production task and the down-time for adaptation of robotic cells or automatic machines to a variation of the production, and increase of the overall line productivity, in terms of overall equipment effectiveness.

\subsection*{Social impact on employment and working conditions}
Some categories of workers are widely recognized as particularly vulnerable, specifically in the presence of a worldwide economic crisis. Old employees, low educated people, and disabled people fall among those categories. The vulnerability of those people is related to the fact that they are the most likely to lose their job, and the less likely to be re-trained and re-employed. This is due either to the difficulty in effectively utilizing complex modern computer aided manufacturing equipment, or to physical impairment that prevents some kinds of activities.

A significant impact on the employment of elderly, low educated and disabled people is expected to result from the application of the proposed methodology. The resulting adaptive HMI, in fact, will automatically adapt to the skills of the current user, supporting the initial (off-line and on-line) training phase, and letting each user reach high levels of productivity in a short time. This will significantly reduce the risk for those people to lose their job due to lack of specific skills. At the same time, in case of loss of the job, it will increase the possibility of re-employment, since the re-training phase is significantly reduced.

Furthermore, as is well known, very often people refuse innovation and automation. Main reasons are related to fear for technological unemployment (i.e. loss of a job due to a technological change) and, more in general, to difficulties in adapting to new technologies and procedures. However, technological innovation is mandatory for achieving the constantly increasing productivity and quality requirements.
The proposed methodology will have a significant impact on the acceptability of automatic machines and robotic cells in traditional production lines. In fact, providing a smart HMI that will adapt to the current user's skill will help in making the users comfortable with innovative technologies and procedures.

We expect that the proposed methodology will have an impact related to employment and working conditions in terms of usability and cognitive load.
Usability will be evaluated based on surveys that will monitor the degree of satisfaction of users comparing traditional HMIs with the smart HMI developed according the proposed approach.
The cognitive load will be computed by non-invasive measurement of different physiological quantities, such as heart rate, blood pressure or pupillary response.

\subsection*{Impact on the market for automatic machines and robotic cells}

As detailed above, we expect a significant impact in the capability, for elderly, low educated or disabled operators, to profitably utilize advanced (and complex) automatic machines and robotic cells. This increased level of acceptability will open new market opportunities, for producers and integrators of automatic machines and robotic cells. In particular, the smart HMI system designed according to the proposed approach will open new market opportunities for automatic machines and robotic cells in traditionally hostile manufacturing environments, such as SMEs and artisan workshops.



\subsection*{Impact on the market for HMI systems}
According to a report published at the end of 2015 \cite{HMI_forecast_2015}, the value of the worldwide HMI market is estimated to reach US\$5,579.3 by 2019, expanding at a CAGR (Compounded Average Growth Rate) of 10.4\% during the period from 2013 to 2019. According to this report, one of the key factors in the growth of HMI market is to be found in the high rate of development in industrial automation: in fact, complex automatic machines and robotic cells require modern HMI systems to be effectively utilized by non-specialized workers in an useful manner.
The market of HMIs is composed of different items: touchscreens or displays, industrial PCs, interface software, and various other controllers. Among these, the market for interface software leads the global HMI market at present: analysts project this market to report the fastest growth during the forecast period.
The application of the proposed methodology will further push this positive trend, as a consequence of the increasing market opportunities for automatic machines and robotic cells.

\section{Conclusions}\label{sec:conclusions}
In this paper, we presented a methodology for the design of adaptive human-centred HMIs for industrial machines and robots. The interfaces developed according to the proposed approach adapt the information presented to the user and its visualization to the user's capabilities and strain level. Thus, they allow for \emph{inclusive} and flexible working environments accessible to any kind of operator, regardless of age, education level, cognitive and physical impairments and experience in the tasks to be performed. Additionally, the proposed approach considers a teaching module that adaptively provides training to unskilled users on the basis of their capabilities and actual understanding of the working scenario.

The approach presented in this paper has been devised within the framework of the European project INCLUSIVE, which is ongoing. Thus, the results of the proposed methodology will be measured at the end of the INCLUSIVE project on three different industrial use cases. For the time being, the expected impact for the increased customization, flexibility optimization of the production and the widest acceptance of automation technologies is investigated in this paper.

\section*{Acknowledgement}
This work has been supported by the INCLUSIVE collaborative project, which has received funding from the European Union's Horizon 2020 Research and Innovation Programme under grant agreement No 723373.

\bibliographystyle{IEEEtran}
\bibliography{ETFA_2017_INCLUSIVE}                     

\end{document}